\documentclass[letter]{aa} 

\usepackage{graphicx}
\usepackage{txfonts}
\usepackage{color}
\usepackage{xspace}
\usepackage{siunitx}
\usepackage{amsmath}
\usepackage{amssymb}
\usepackage{xcolor}
\usepackage{dashbox}
\usepackage{framed}
\usepackage{lipsum}
\usepackage{placeins}
\usepackage{stfloats}

\usepackage{hyperref}

\hypersetup{
        colorlinks=true,                                                
        breaklinks=true,
        linkcolor=blue,                                                     
        citecolor=blue,                                                         
        filecolor=blue,                                                     
        urlcolor=blue,
        unicode=false,                                                      
        pdftoolbar=true,                                                    
        pdfmenubar=true,                                                    
        pdffitwindow=false,                                                 
        pdfstartview={Fit},                                                 
        pdftitle={Extended stellar systems in the solar neighborhood II},  
        pdfauthor={Stefan Meingast},                        
        pdfsubject={},                                          
        pdfcreator={Stefan Meingast},                       
        pdfkeywords={}, 
        pdfnewwindow=true,                                                  
        pdfdisplaydoctitle=true                                     
}

\makeatletter
\renewcommand*\aa@pageof{, page \thepage{} of \pageref*{LastPage}}
\makeatother

\definecolor{stefan}{rgb}{0.86, 0.08, 0.24}

\definecolor{joao}{rgb}{0.01, 0.75, 0.24}

\definecolor{verena}{rgb}{0.13, 0.55, 0.13}

\newcommand{\code}{\texttt}
\newcommand{\gaia}{\textit{Gaia}\xspace}

\begin{document}

\defcitealias{meingast18}{Paper~I}

\title{Extended stellar systems in the solar neighborhood}
\subtitle{II. Discovery of a nearby $120^{\circ}$ stellar stream in \gaia DR2}


\author{Stefan Meingast\inst{1}
        \and Jo\~ao Alves\inst{1,2,3}
        \and Verena F\"urnkranz\inst{1}
        }
            
\institute{Department of Astrophysics, University of Vienna, T\"urkenschanzstrasse 17, 1180 Wien, Austria
\\ \email{stefan.meingast@univie.ac.at}
\and
Radcliffe Institute for Advanced Study, Harvard University, 10 Garden Street, Cambridge, MA 02138, USA
\and
Data Science @ Uni Vienna, Faculty of Earth Sciences Geography and Astronomy, University of Vienna, Austria
}

\date{Received 21 December 2018 / Accepted 17 January 2019}

\abstract{We report the discovery of a large, dynamically cold, coeval stellar stream that is currently traversing the immediate solar neighborhood at a distance of only \SI{100}{pc}. The structure was identified in a wavelet decomposition of the 3D velocity space of all stars within \SI{300}{pc} to the Sun. Its members form a highly elongated structure with a length of at least \SI{400}{pc}, while its vertical extent measures only about \SI{50}{pc}. Stars in the stream are not isotropically distributed but instead form two parallel lanes with individual local overdensities, that may correspond to a remnant core of a tidally disrupted cluster or OB association. Its members follow a very well-defined main sequence in the observational Hertzsprung-Russel diagram and also show a remarkably low 3D velocity dispersion of only \SI{1.3}{\km \per \s}. These findings strongly suggest a common origin as a single coeval stellar population. An extrapolation of the present-day mass function indicates a total mass of at least \SI{2000}{M_\odot}, making it larger than most currently known clusters or associations in the solar neighborhood. We estimated the stream’s age to be around \SI{1}{Gyr} based on a comparison with a set of isochrones and giant stars in our member selection and find a mean metallicity of $\left[ \mathrm{Fe/H} \right] = -0.04$. This structure may very well represent the Galactic disk counterpart to the prominent stellar streams observed in the Milky Way halo. As such, it constitutes a new valuable probe to constrain the Galaxy's mass distribution.
}


\maketitle

\section{Introduction}
\label{sec:introduction}

The dynamic evolution of stellar systems is largely governed by the gravitational potential of their host galaxies. Given enough time, star clusters generally seem to be shaped into dynamically cold, distinctly elongated structures \citep[e.g.][]{Grillmair95, Odenkirchen01, Grillmair06}. So far, extended stellar streams have only been observed for tidally disrupting clusters that are associated with the Milky Way halo \citep[e.g.][]{malhan18}. Since their discovery, these systems have proven to be one of the most promising probes for constraining the mass distribution of the Galaxy on both small and large scales, including its dark matter component \citep[e.g.][]{Bovy2016,Ibata2016,Price-Whelan18}.

This paper series is dedicated to studying such morphological characteristics of dynamically evolved stellar system in the immediate solar neighborhood. As their halo counterparts, Galactic disk streams could be used as additional tools for measuring the Milky Way mass distribution and could bring unique insight into the formation of the Galactic field through dissolution of clusters and associations. In the previously published letter \citep[][hereinafter Paper I]{meingast18}, we demonstrated that the nearest open cluster, the Hyades, show well-developed tidal tails, largely in agreement to theoretical predictions (see also \citealp{roeser18}). In this manuscript, we extend our investigation beyond currently known clusters in the solar neighborhood. Specifically, we identified and characterized a large, coeval stellar stream that is tied to the Galactic disk and is located at a distance of only \SI{100}{pc} to the Sun.

\section{Data}
\label{sec:data}

As in \citetalias{meingast18}, we used the 6D position-velocity information provided with the second \gaia data release \citep[\gaia DR2;][]{gaia_mission, gaia_dr2}. Distances were obtained via the catalog published by \citet{bailer-jones18}. For this letter, however, we constructed different subsets from the initial DR2 database when compared to the previously published paper in this series. Specifically, we increased the distance limit to $d \leq \SI{300}{pc}$, and at the same time relaxed our error criteria on data quality to $\sigma_{\varpi} / \varpi < 0.5$, $\sigma_{\mu_{\alpha,\delta}} / \mu_{\alpha,\delta} < 0.5$, $\sigma_{rv} / rv < 0.5$, and $\sigma_{G_{BP,RP}} / G_{BP,RP} < 0.5$. All other criteria and definitions (e.g. the coordinate system used) remained the same as in \citetalias{meingast18}.

Testing a series of different quality criteria showed that our results did not change significantly even without any restrictions on measurement errors. This insensitivity to errors is most likely caused by the subsequently applied clustering algorithm (see Sect. \ref{sec:id}) which requires a connection between sources in the 6D position-velocity space. Nevertheless, we chose to retain the error quality criteria, motivated by our preference to publish a reliable final sample. With the above-listed filtering setup, we find a total number of \SI{749611}{} sources in the \gaia database.

\begin{figure*}
        \centering
        \resizebox{1.0\hsize}{!}{\includegraphics[]{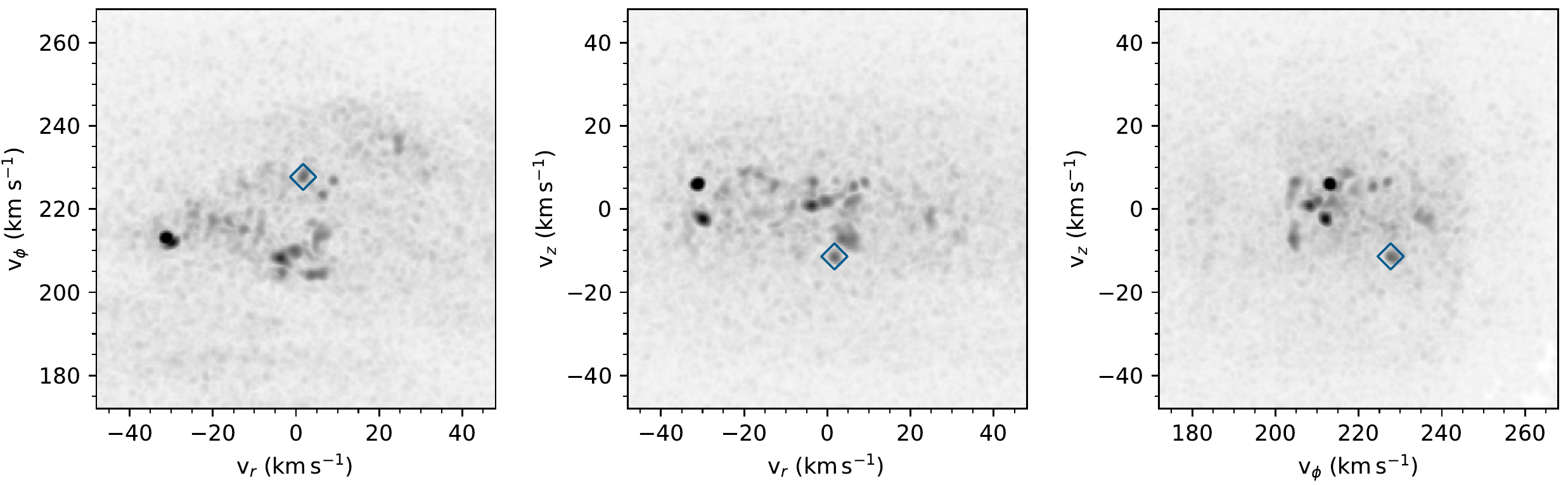}}
        \caption[]{Wavelet component, tracing scales on the order of \SI{1.5}{\km \per \s} in 3D velocity space. The panels show the distribution from different sides where the greyscale colormap is proportional to number density and shows the maximum value along the viewing direction. The blue diamond marks the velocity of the stream discussed in this letter.}
    \label{img:wd}
\end{figure*}

\section{Stream identification}
\label{sec:id}

We started with the assumption that dynamically cold stellar populations can be identified as overdensities in velocity space. Following \citetalias{meingast18}, we used a cylindrical coordinate system ($r, \phi, z$) with its origin at the Galactic center (the corresponding velocities are $v_r, v_{\phi}, v_z$). To obtain density information in velocity space, we initially constructed a 3D histogram of the cylindrical velocity components. 
Here, we limited the histogram size to $v_r, v_z \in \left[-48, 48 \right]$ \SI{}{\km \per \s} and $v_{\phi} \in \left[ 172, 268 \right]$ \SI{}{\km \per \s} and chose a bin-size of \SI{0.375}{\km \per \s}. These specific limits were motivated by a restriction in the subsequently applied wavelet decomposition with the \code{PyWavelets} \code{Python} package \citep{pywavelets}, which requires the initial array (the histogram) to have an edge length drawn from the set $\left\{ 2^x\; \forall\; x \in \mathbb{N} \right\}$ for optimal performance. 
Briefly explained, the wavelet decomposition constructs a series of components, each representing different scales (in our case density in velocity space). In Fig.~\ref{img:wd} we show the component tracing scales on the order of \SI{1.5}{\km \per \s}, which is a typical value for the velocity dispersion of young associations and moving groups \citep[e.g.,][and references therein]{Preibisch08,riedel17}. In this figure, the three panels are views of the 3D array from different sides, where the displayed grayscale map represents the maximum array value along the view direction.

Figure~\ref{img:wd} reveals a complex structure in velocity space with many individual overdensities for which we performed an extraction of significant peaks by applying a 5-$\sigma$ threshold above the mean level of the array. Most of the peaks obtained in this way could be associated to already known clusters, associations, or moving groups. 
However, we identified a very prominent, previously unknown structure at $v_r=$\;\SI{1.7}{\km \per \s}, $v_{\phi}=$\;\SI{227.7}{\km \per \s}, $v_z=$\;\SI{-11.4}{\km \per \s} (marked in Fig. \ref{img:wd} with a blue diamond symbol) which is the main subject of this letter. In light of the complexity and information density in velocity space, a full extraction and analysis of significant overdensities will be published in the near future as part of this paper series.

Based on the methods used in \citetalias{meingast18}, we first extracted all sources from our \gaia database within a \SI{5}{\km \per \s} radius around the identified velocity peak, leaving \SI{2160}{} sources. We then continued to search for relations between individual stars in the 6D parameter space spanned by $r, \phi, z, v_r, v_{\phi}, v_z$. As the velocities and spatial coordinates probe different value ranges, we first pre-processed the data space so that each parameter featured a mean of 0 and unit variance. 
Within this scaled parameter space, we then searched for clusters with the popular clustering algorithm DBSCAN \citep{dbscan}. 
We performed a series of iterations with variable setups for the clustering algorithm, where the quality of the extraction was judged by investigating the resulting observational Hertzsprung-Russel diagrams (HRD), as well as the spatial source distribution. Almost all setups returned a single very prominent group for which our final DBSCAN setup for the scaled parameters ($\mathrm{minPts}=50$, $\epsilon=0.75$) was chosen as a compromise between the number of extracted sources (\SI{440}{}) and a relatively clean HRD. Where applicable, sources belonging to this subset are displayed as small blue dots in the figures in this manuscript.

\begin{figure*}[t!]
        \centering
        \resizebox{\hsize}{!}{\includegraphics[]{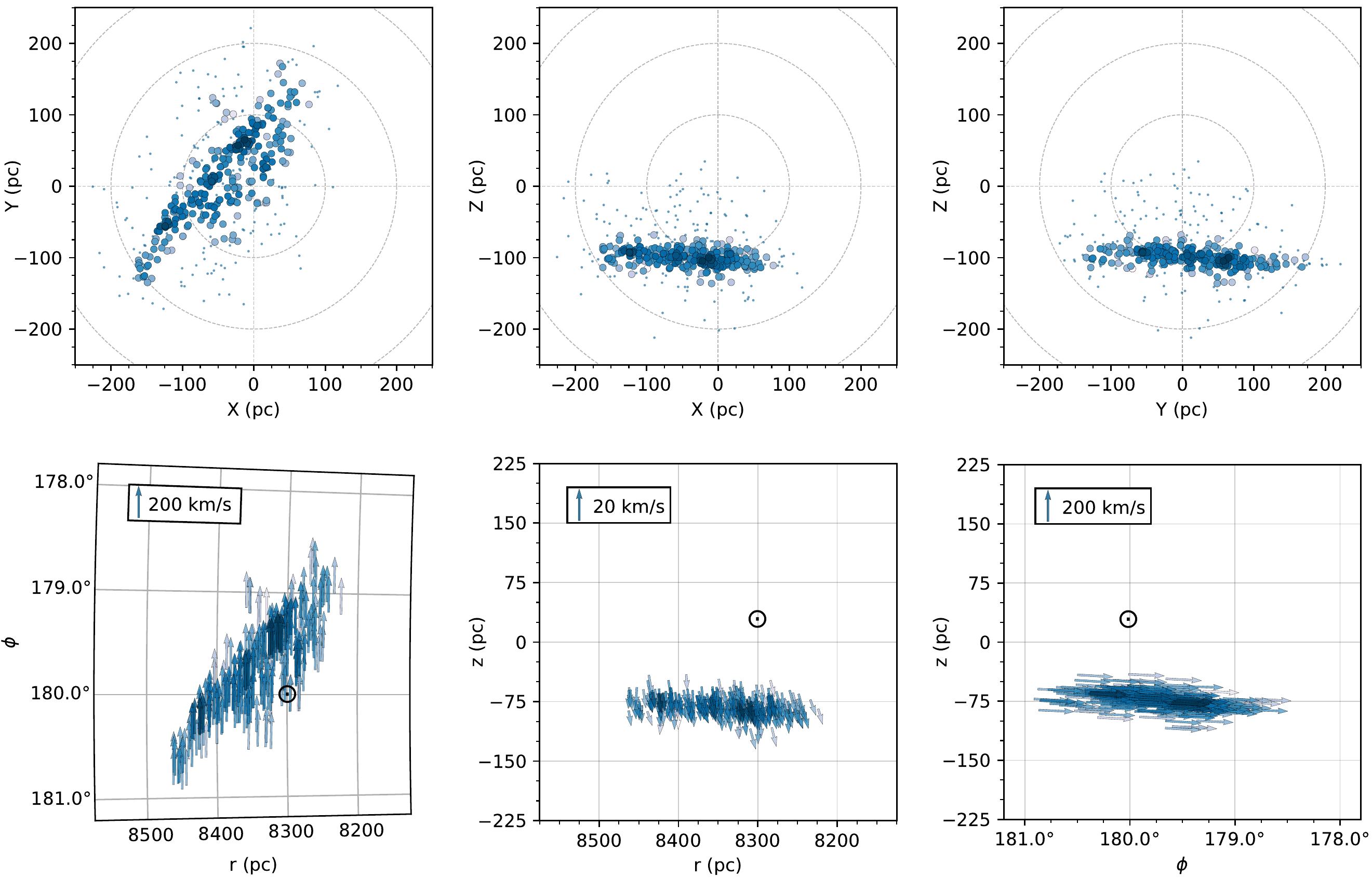}}
        \caption[]{Positions of our final stream member selection in Galactic Cartesian (top, centered on the Sun) and Galactocentric cylindrical (bottom) coordinates. In the Cartesian frame X is positive toward the Galactic center, Y is positive in the direction of Galactic rotation, and Z points to the Galactic north pole. In the bottom row, the Sun's position is indicated with the black circular symbol. The small blue dots correspond to all sources that were identified in our 6D clustering application, while the filled circles are sources in our final stream selection. The colorscale of the filled circles is proportional to source density, with darker shades referring to higher densities. The arrows in the bottom row visualize the velocity vector. The sources depict a strikingly flat spatial arrangement with an extent of at least \SI{400}{pc} along its major axis.}
    \label{img:xyz}
\end{figure*}

Inspired by the selection criteria in \citetalias{meingast18}, we subsequently applied a spatial density threshold. Also here, we tested several setups and decided for a final criterion that selected only sources with at least 7 neighbors within a radius of \SI{30}{pc}. This criterion reduced the final sample to \SI{258}{} sources. Their distribution on the HRD revealed two sources well below the main sequence for which also typical \gaia quality control parameters did not show suspicious values. We removed these two sources manually, resulting in a final selection of \SI{256}{} sources\footnote{A subset of this selection, along with computed velocities, is listed in Table~\ref{tab:sources}, while the full catalog is available online via CDS.}. Figure~\ref{img:xyz} displays the source positions for this final selection in both Galactic Cartesian coordinates and Galactocentric cylindrical coordinates. Figure~\ref{img:allsky} shows the distribution on sky for Equatorial and Galactic coordinates where we find an extent of \SI{120}{\degree}.

As suggested by the referee, we estimated the contamination fraction of our selection by extracting sources in a symmetric phase-space region on the opposite side of the Galactic plane. First, we obtained all Gaia sources (same error cuts) within a \SI{4}{\km \per \s} radius (3-$\sigma$ of the structure's velocity dispersion; Sect.~\ref{sec:structure}) around the stream's velocity coordinates with inverted vertical velocity (i.e. $v_z = \SI{11.4}{\km \per \s}$). Then, we extracted all sources in this subset that fall into a mirrored volume with respect to the Galactic plane. Specifically, we achieved this by projecting each stream source on the other side of the Galactic plane and only retaining sources in the new selection that have a mirrored stream neighbor within \SI{13}{pc}. This value was chosen, because the maximum nearest neighbor distance for our stream sources is \SI{26}{pc}. In this way, we obtained a selection for which the orbit energies and the z-components of the angular momenta would be similar to the sources in the stream. Without any further restrictions to spatial density or subsequent DBSCAN clustering, this set contained only four sources, indicating a contamination level of a few percent.

\begin{figure}
        \centering
        \resizebox{1.0\hsize}{!}{\includegraphics[]{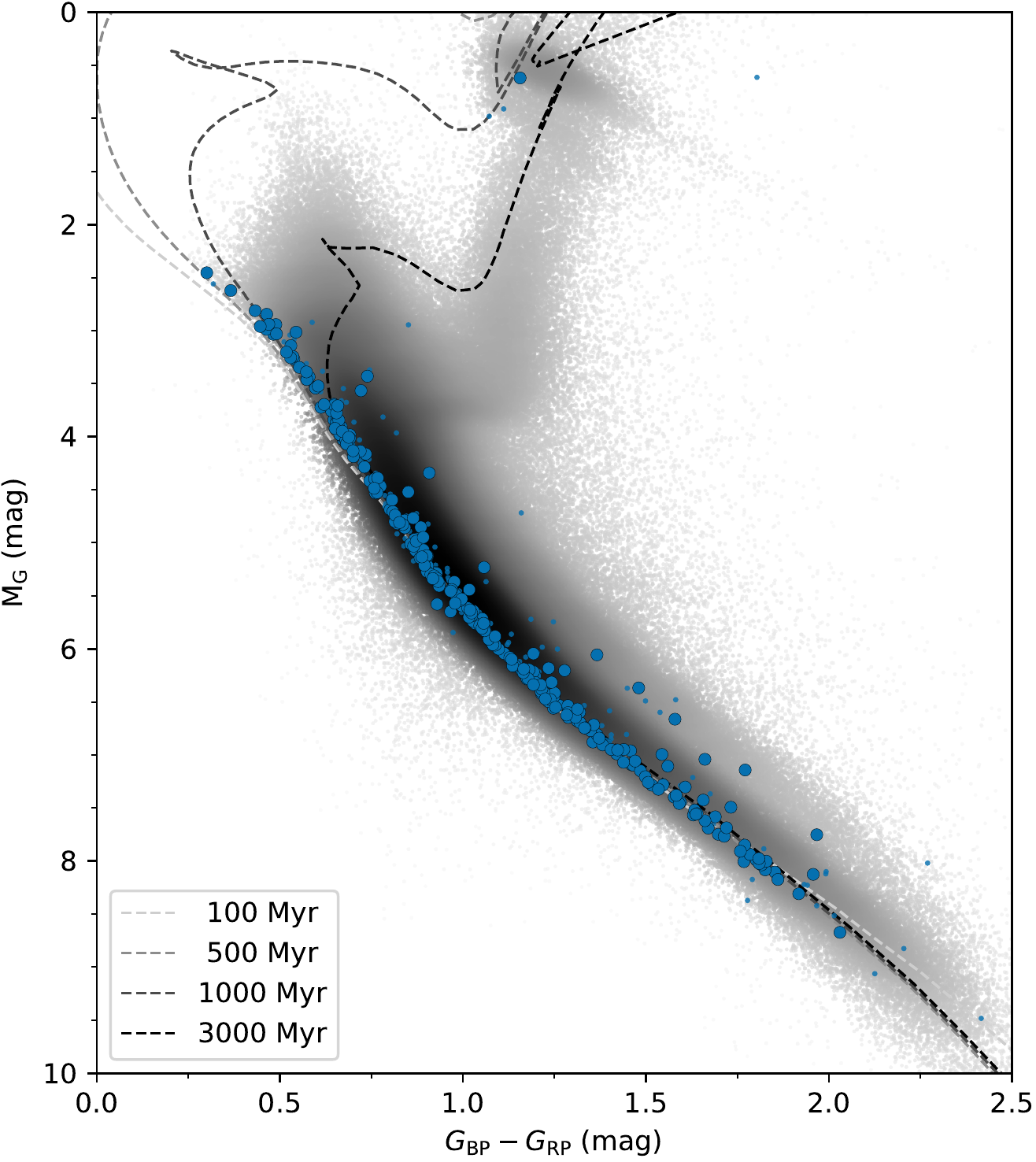}}
        \caption[]{Observational HRD for stream members, constructed with photometry from \gaia DR2. The filled circles and small dots are the same as in Fig~\ref{img:xyz}. The gray dots in the background are all sources with radial velocity measurements within \SI{300}{pc}. The dashed lines show a set of isochrones with their age labeled in the bottom left corner.}
    \label{img:hrd}
\end{figure}

The selected sources form a striking, highly elongated, stream-like structure that currently traverses the solar vicinity only about \SI{{\sim}100}{pc} below the Sun's current position toward the Galactic south pole (\SI{{\sim}75}{pc} below the Galactic midplane). In anticipation of the following analysis, Fig.~\ref{img:hrd} also reveals a remarkably well-defined main sequence, strongly suggesting a coeval stellar population. We repeated the 2D KS test on the HRD used in \citetalias{meingast18} and found a p-value of $5 \times 10^{-4}$, confirming its coeval nature.

\section{Results and discussion}
\label{sec:discussion}


\subsection{Structure and kinematics}
\label{sec:structure}

The stream extends for a length of about \SI{400}{pc} with a vertical size of only about \SI{50}{pc}. We additionally measured an axial ratio of 8.4:2.7:1.0 by means of the $XYZ$ covariance. We note here that due to the sensitivity limit of \gaia's spectrometer, the structure most likely contains sources beyond the current distance limitations and therefore is potentially much larger. Also, similar to the Hyades cluster analyzed in \citetalias{meingast18}, the structure is elongated with respect to its direction of movement (lower right panel of Fig.~\ref{img:xyz}). In terms of the stream's kinematics, we measure an astonishingly low 3D velocity dispersion of only \SI{1.3}{\km \per \s} across the entire structure. While its current vertical velocity of \SI{-11.4}{\km \per \s} clearly shows that it is still moving away from the Galactic plane, the velocity is not high enough for the stream to escape into the Milky Way halo.

The general structure, kinematics, and the coeval nature are distinct characteristics that can also be attributed to tidally disrupted clusters observed in the Milky Way halo \citep{malhan18}. These findings therefore suggest that the observed stream is in fact either a fully disrupted cluster or association, or a system that is still continuing to be disrupted.


Inspecting the source distribution in Fig.~\ref{img:xyz} more closely reveals additional interesting attributes. First, the source density along the structure is not isotropic. To highlight the differences, we color-coded each source in this figure by its density (darker shades of blue refer to higher density). Moreover, Fig.~\ref{img:kde_xyz} in the Appendix shows kernel-density maps constructed with a \SI{20}{pc} wide parabolic (Epanechnikov) kernel on an 8-fold oversampled grid for better visualization. Intriguingly, the source distribution shows individual clumps in density. We tested whether this distinct arrangement could be an artifact caused by the inhomogeneous sky coverage of \gaia DR2, but the overdensities did not show an obvious correlation with \gaia coverage maps. We determined the 3D positions of the overdensities by calculating a weighted mean of the source positions in a \SI{20}{pc} radius around a manually selected center. The results are marked with arrows in Fig.~\ref{img:kde_xyz} and are listed in Table~\ref{tab:knots}, where the labels are ordered by peak density (label 1 is the highest density). There are no obvious matches between already known moving groups or clusters and the given coordinates. We speculate that these overdensities could correspond to remnants of stellar clusters that are currently in their last phases of disruption. 

Secondly, Figs.~\ref{img:xyz} and \ref{img:kde_xyz} seem to split the structure into two parallel streams. This particular attribute is reminiscent of the spur feature observed for the prominent globular cluster stream GD-1 \citep{Price-Whelan18, Bonaca18}. A split by hand revealed that the most significant difference between the two subsamples can be found in their $v_\phi$ velocity component, where the sub-stream closer to the Galactic center currently moves on average faster by \SI{1}{\km \per \s}. 

\subsection{Age}
\label{sec:age}

Figure~\ref{img:hrd} shows the observational HRD along with a set of PARSEC isochrones \citep{Bressan12} for solar metallicity. While no clear main sequence turnoff is apparent, a single source in our main selection is located close to the red clump. That source is 42 Ceti, a multiple star where one component is classified as a G8IV subgiant \citep{houk99}. There are two more sources located near the red clump that were removed by the spatial density filter: HD 57727 (57 Gem) and HD 203382. Both of these are G8III giants \citep{harlan69}. Assuming these are part of the stream, a comparison of the isochrones and the three giant stars suggests and age of around \SI{1}{Gyr}.

Another argument in favor of such an advanced age can be made by a comparison to the Hyades. The Hyades' age is estimated to be about \SI{600}{Myr} \citep{perryman98}, possibly even \SI{800}{Myr} \citep{brandt15}. According to \citet{ernst11}, the initial mass of the Hyades was about \SI{1700}{M_\odot}. In Sect.~\ref{sec:mass} we argue that the structure must have had an initial mass above \SI{2000}{M_\odot}, thus exceeding the initial mass for the Hyades. However, in contrast to the newly discovered stream, the Hyades still have a pronounced core. Assuming the evolution of the two groups is comparable (e.g. no fatal encounters with giant molecular clouds), the new stream must therefore be dynamically older.

\subsection{Total mass and metallicity}
\label{sec:mass}

A direct mass estimate for the entire structure is aggravated by several biases in the source selection. Most notably, \gaia delivers radial velocities (which are a prerequisite for our selection) only for a certain range of spectral types. Also, as the structure spans several hundred pc across \SI{120}{\degree} on sky, sensitivity limitations are also a function of position. Moreover, mass segregation along the stream and stellar evolution can influence the observed population. The following mass estimate is based on extrapolated initial mass functions (IMF) and should be considered a lower limit for the stream's total mass.

\begin{figure}
        \centering
        \resizebox{1.0\hsize}{!}{\includegraphics[]{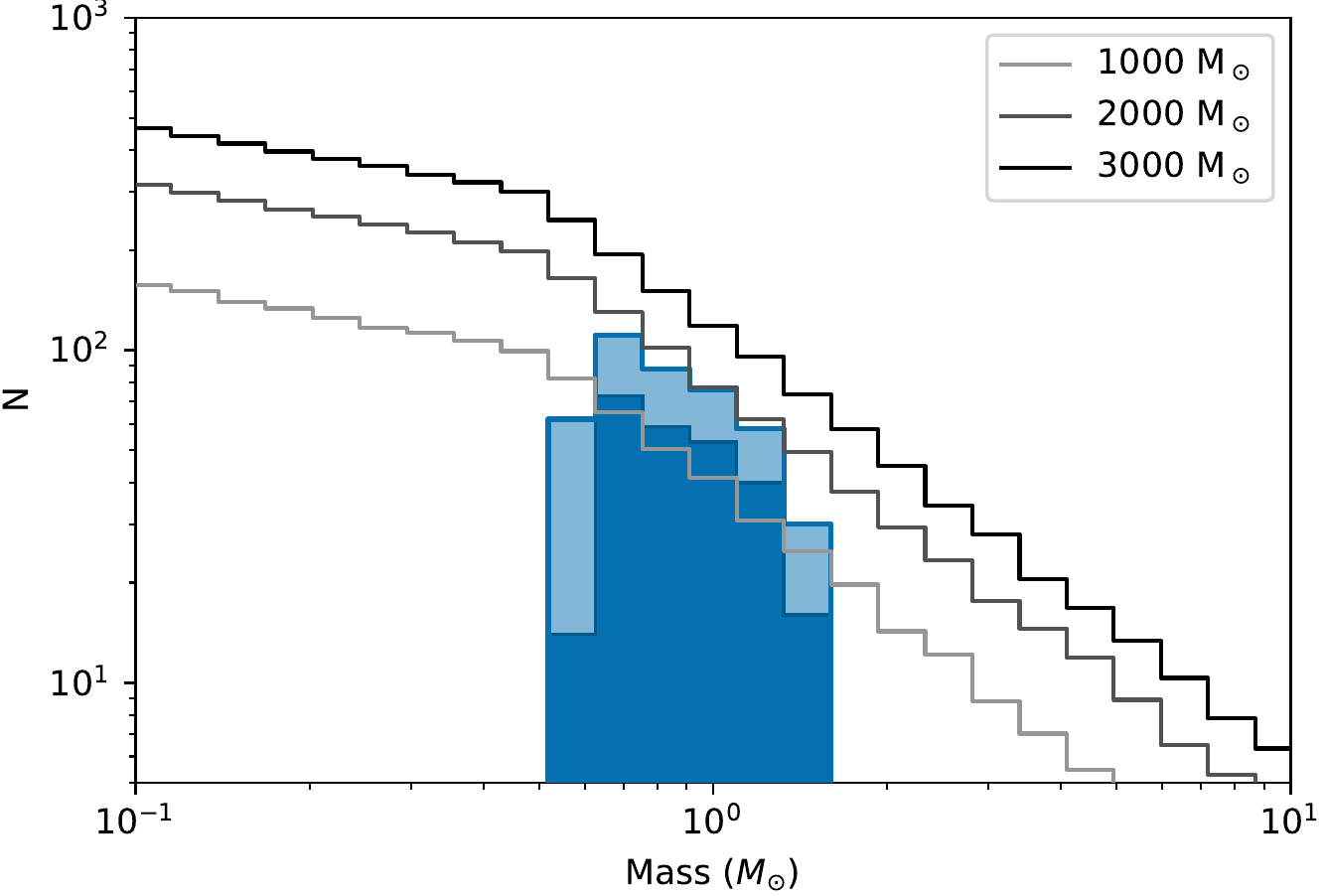}}
        \caption[]{Present-day mass function for all selected stream members. While the dark blue histogram shows our selected members, the pale blue histogram incorporates a coarse completeness correction. On top, we show a series of initial mass functions with their total cluster masses labeled in the top right corner.}
    \label{img:mf}
\end{figure}

Figure~\ref{img:mf} shows the mass distribution for all sources in our selection for which individual masses were estimated by interpolating a \SI{1}{Gyr} PARSEC isochrone with solar metallicity (dark blue histogram). Also shown are IMFs \citep[from][]{kroupa01} for \SI{1000}{M_\odot}, \SI{2000}{M_\odot}, and \SI{3000}{M_\odot} clusters. As our selection is affected by incompleteness, we applied a correction assuming a similar bias as found in \citetalias{meingast18}. There, about 70\% of the Hyades members are missing in the \gaia selection within the observed mass range. We added these missing sources to our selection based on the given IMF probability distribution. The resulting histogram is displayed in pale blue. The completeness-corrected mass distribution shows a reasonably good match with a \SI{2000}{M_\odot} IMF. For comparison, the nearest massive star-forming site to the Sun, the Orion Nebula Cluster, is estimated to hold about \SI{1000}{M_\odot} \citep{dario12}. Thus, the event that produced this structure may have been similar to the most massive clusters we currently observe in the Galaxy. 

In terms of stellar mass, the newly discovered structure is similar to the recently found Phlegethon stream \citep{Ibata18}. However, Phlegethon, presumably a \SI{10}{Gyr} old remnant of a disrupted globular cluster, is characterized by very low metallicity of $\left[ \mathrm{Fe/H} \right] \approx -1.5$. In contrast, a cross-match (\SI{1}{arcsec} radius) of our sources with LAMOST DR4 \citep{Cui12} yields 27 matches with a mean metallicity of $\left[ \mathrm{Fe/H} \right] = -0.04 \pm 0.15$. The evident dissimilarities with respect to age and metallicity clearly point to an entirely different origin of the two systems.

\section{Summary and conclusions}
\label{sec:summary}

Using kinematic information provided with \gaia DR2, we identified a remarkably large and dynamically cold stream of stars in the immediate solar vicinity. It extends for at least \SI{400}{pc} along its major axis while being vertically much more confined. The stream is currently located about \SI{75}{pc} below the Galactic midplane, where it still continues to move away from the Galactic plane (Fig.~\ref{img:xyz}). Its vertical velocity, however, is not high enough for the stream to escape into the halo. Furthermore, it is highly elongated along its direction of movement and shows an exceptionally low 3D velocity dispersion of only \SI{1.3}{\km \per \s} across the entire structure. Moreover, the member stars form a very clear main sequence in the HRD, strongly suggesting a coeval nature (Fig.~\ref{img:hrd}). Based on a set of isochrones and a comparison to the Hyades, we find an age of about \SI{1}{Gyr}. We estimate its minimum total mass to be around \SI{2000}{M_\odot}, making it substantially larger than most other nearby evolved moving groups or clusters (Fig.~\ref{img:mf}). Furthermore, we determine a mean metallicity of $\left[ \mathrm{Fe/H} \right] = -0.04 \pm 0.15$ via cross-matching to the LAMOST survey.

All presented evidence taken together strongly suggests that this newly discovered stream is a direct counterpart to stellar streams observed in the Milky Way halo. It appears that also clusters and associations bound to the disk are gradually shaped into thin, dynamically cold stellar streams. Barring fatal encounters with molecular clouds, these can orbit the Galaxy for several hundred Myrs, before eventually dissolving into the Galactic field population. Similar to their halo analogues, such streams can act as important test cases for the large and small scale mass distribution of the Galaxy.

\begin{acknowledgements}
We thank the referee, Rodrigo Ibata, for carefully reading this manuscript and for his useful comments that served to improve both the clarity and quality of this study.
This work has made use of data from the European Space Agency (ESA) mission \gaia (\url{https://www.cosmos.esa.int/gaia}), processed by the \gaia Data Processing and Analysis Consortium (DPAC, \url{https://www.cosmos.esa.int/web/gaia/dpac/consortium}). Funding for the DPAC has been provided by national institutions, in particular the institutions participating in the \gaia Multilateral Agreement.
This research made use of Astropy, a community-developed core Python package for Astronomy \citep{astropy}.
This research has made use of "Aladin sky atlas" developed at CDS, Strasbourg Observatory, France \citep{bonnarel00}.
We also acknowledge the various Python packages that were used in the data analysis of this work, including NumPy \citep{numpy}, SciPy \citep{scipy}, scikit-learn \citep{scikit-learn}, scikit-image \citep{scikit-image}, and Matplotlib \citep{matplotlib}.
This research has made use of the SIMBAD database operated at CDS, Strasbourg, France \citep{simbad}.
\end{acknowledgements}

\bibliography{references}
\clearpage

\begin{appendix}

\section{Supplementary plots and tables}
\label{sec:app}

\begin{figure*}[t]
    \centering
    \resizebox{\hsize}{!}{\includegraphics[]{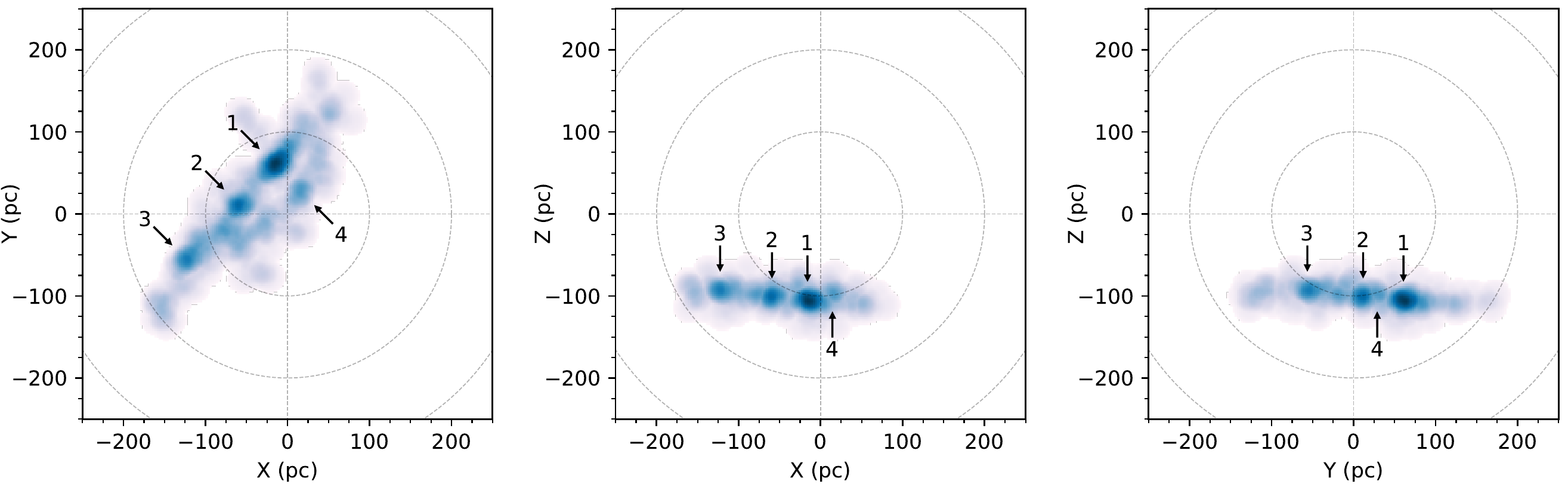}}
    \caption[]{Kernel density maps for our stream member selection in Galactic Cartesian coordinates, using a \SI{20}{pc} Epanechnikov kernel. The stream clearly shows local overdensities and also appears to be split into two parallel lanes. The labels are referencing the IDs as listed in Table~\ref{tab:knots}.}
    \label{img:kde_xyz}

    \vspace*{\floatsep}

    \centering
    \resizebox{\hsize}{!}{\includegraphics[]{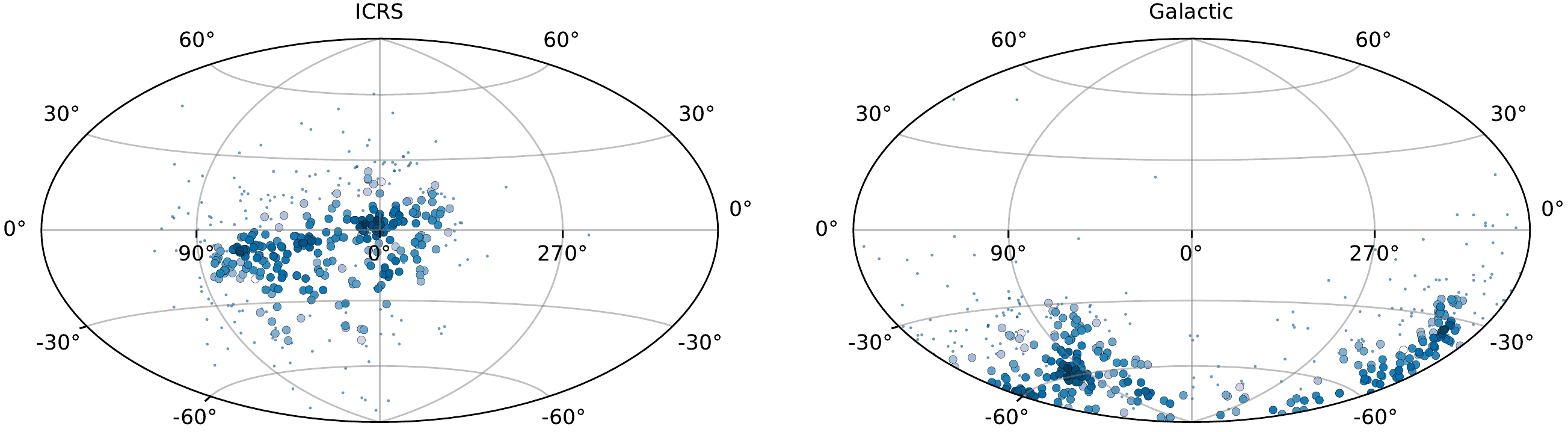}}
    \caption[]{Allsky distribution of our member selection. The left-hand side panel displays the sources in Equatorial coordinates (ICRS), the right-hand side in Galactic coordinates (right). The data points are the same as in Fig.\ref{img:xyz} which reveal an extent of about \SI{120}{\degree}.}
    \label{img:allsky}
\end{figure*}

Here we provide additional information that is referenced in the main text. Figure~\ref{img:kde_xyz} shows kernel density maps in Galactic Cartesian coordinates and Fig.~\ref{img:allsky} shows an allsky distribution of our member selection that is listed in Table~\ref{tab:sources}. The former figure reveals an inhomogeneous source distribution, with individual labeled overdensities. For these, we list their coordinates in Table~\ref{tab:knots}.

\begin{table*}
    \begin{tabular*}{\linewidth}{l @{\extracolsep{\fill}} c c c c c c c c}
        \hline\hline
Gaia DR2 source\_id	&	ra	&	dec	&	$r$	&	$\phi$	&	$z$	&	$v_r$	&	$v_\phi$	&	$v_z$	\\
	&	(deg)	&	(deg)	&	(pc)	&	(rad)	&	(pc)	&	(km/s)	&	(km/s)	&	(km/s)	\\
\hline																	
7324465427953664	&	46.30862	&	6.14818	&	8403.36	&	3.13989	&	-71.72	&	-1.24	&	226.96	&	-12.27	\\
24667616384335872	&	36.94767	&	11.24092	&	8386.11	&	3.13735	&	-65.83	&	-1.95	&	226.76	&	-11.25	\\
1776592713522549376	&	332.25859	&	18.95330	&	8265.43	&	3.12078	&	-72.26	&	-2.38	&	228.35	&	-12.28	\\
2326464319728427264	&	356.25468	&	-31.53067	&	8276.71	&	3.14099	&	-62.18	&	-3.17	&	230.49	&	-11.71	\\
2336552648151249280	&	0.76617	&	-24.84388	&	8286.97	&	3.14014	&	-64.31	&	-2.32	&	231.54	&	-14.28	\\
2339984636258635136	&	359.22381	&	-23.29018	&	8287.00	&	3.13987	&	-58.11	&	-1.80	&	230.02	&	-10.77	\\
2346216668164370432	&	13.55613	&	-22.88550	&	8305.20	&	3.14092	&	-71.94	&	-2.00	&	229.27	&	-12.80	\\
2349094158814399104	&	11.82496	&	-22.75224	&	8302.85	&	3.14076	&	-67.00	&	-1.89	&	229.20	&	-9.80	\\
2355466790769878400	&	13.83938	&	-21.40102	&	8306.29	&	3.14078	&	-61.97	&	0.31	&	229.74	&	-9.94	\\
2358736291674305408	&	18.56869	&	-16.06257	&	8323.26	&	3.13998	&	-94.91	&	-0.46	&	227.44	&	-12.18	\\
\hline
    \end{tabular*}
    \caption{Subsample of our selected stream members. The full selection, including additional columns (positions and velocities in the Galactic Cartesian frame) is available online via CDS.}
    \label{tab:sources}
\end{table*}

\begin{table}[!hb]
    \begin{tabular*}{\linewidth}{l c c  @{\extracolsep{\fill}} c c c c}
	\hline\hline
	ID              &  $l$ & $b$ & $X$ & $Y$ & $Z$ & $d$  \\
	                 & (deg) & (deg)  & (pc) & (pc) & (pc) & (pc) \\
    \hline
    1            & 204.74 & -34.51   & -122.5 & -56.4 & -92.7  & 163.7  \\
    2            & 168.96 & -59.09   & -59.3  & 11.6  & -100.9 & 117.6  \\
    3            & 104.69 & -59.12   & -15.9  & 60.7  & -105.0 & 122.3  \\
    4            & 63.31  & -71.55   & 14.5   & 28.8  & -96.5  & 101.7  \\
    \hline
	\end{tabular*}
    \caption{Coordinates of the overdensities highlighted in Fig.~\ref{img:kde_xyz}. The parameters $l$ and $b$ refer to Galactic sky coordinates, while $XYZ$ denote Galactic Cartesian coordinates centered on the Sun. The distance is given in the column labeled $d$.}
	\label{tab:knots}
\end{table}

\end{appendix}

\end{document}